\documentclass[letter]{aa}
\usepackage{txfonts}
\usepackage{graphicx}
\begin{document}

\title{The close-in companion of the fast rotating Be star Achernar\thanks{Based on observations collected at the European Southern Observatory, Chile under ESO Program 279.D-5064(A)}}
\titlerunning{The close-in companion of the fast rotating star Achernar}
\authorrunning{P. Kervella et al.}
\author{
P. Kervella\inst{1}
\and
A. Domiciano~de~Souza\inst{2}
\and
Ph. Bendjoya\inst{2}
}
\offprints{P. Kervella}
\mail{Pierre.Kervella@obspm.fr}
\institute{
LESIA, Observatoire de Paris, CNRS\,UMR\,8109, UPMC, Universit\'e Paris Diderot, 5 Place Jules Janssen, 92195 Meudon, France
\and
Lab. H. Fizeau, CNRS UMR 6525, Univ. de Nice-Sophia Antipolis, Observatoire de la C\^ote dÕAzur, 06108 Nice Cedex 2, France
}
\date{Received ; Accepted}
\abstract
{The Be stars are massive dwarf or subgiant stars that present temporary emission lines in their spectrum, and particularly in the $H{\alpha}$ line. The mechanism triggering these Be episodes is currently unknown, but binarity could play an important role.}
{Previous observations with the VLT/VISIR instrument (Kervella \& Domiciano de Souza~2007) revealed a faint companion to Achernar, the brightest Be star in the sky. The present observations are intended to characterize the physical nature of this object.}
{We obtained near-IR images and an H-band spectrum of Achernar B using the VLT/NACO adaptive optics systems.}
{Our images clearly show the displacement of Achernar B over a portion of its orbit around Achernar A. Although there are not enough data to derive the orbital parameters, they indicate a period of about 15\,yr. The projected angular separation of the two objects in December 2007 was less than 0.15$\arcsec$, or 6.7\,AU at the distance of Achernar.}
{From its flux distribution in the near- and thermal-infared, Achernar B is most likely an A1V-A3V star. Its orbital period appears similar to the observed pseudo-periodicity of the Be phenomenon of Achernar. This indicates that an interaction between A and B at periastron could be the trigger of the Be episodes.}

\maketitle

\section{Introduction}

As the brightest ($m_V=0.46$) and nearest Be star in the sky, Achernar ($\alpha$\,Eri, \object{HD\,10144}) has been the focus of a lot of interest over the past decades. Its very fast rotation velocity $v \sin i$ is estimated between 220 to 270\,km.s$^{-1}$ and its effective temperature between $15\,000$ to $20\,000$\,K (Vinicius et al.~\cite{vinicius06}). Achernar was chosen as the subject of the first VLTI observations, which revealed its extraordinarily distorted interferometric profile (Domiciano de Souza et al.~\cite{domiciano03}).  Different possibilities have been proposed recently to explain the exceptionally high flattening ratio of the photosphere of the star (Jackson et al.~\cite{jackson04}; Carciofi et al.~\cite{carciofi08}). Further interferometric observations have revealed the presence of the stellar wind emitted by the overheated poles of the star (Kervella \& Domiciano de Souza~\cite{kervella06}), resulting in a slight revision of its flattening ratio. A model of the envelope of Achernar has recently been presented by Meilland~(\cite{meilland07}).
Last year, we discovered a close-in faint companion to Achernar, from diffraction-limited thermal IR imaging with VLT/VISIR (Kervella \& Domiciano de Souza~\cite{kervella07}). The present Letter reports the follow-up adaptive optics observations in the near-IR domain to characterize this companion, hereafter referred to as Achernar~B.

\section{Observations \label{observations}}

\subsection{Imaging}

\begin{figure}[]
\centering
\includegraphics[width=4.4cm]{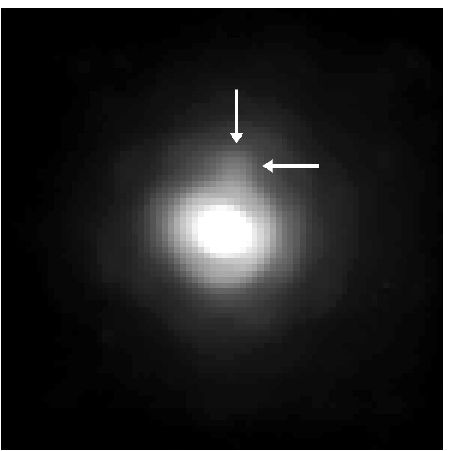}
\includegraphics[width=4.4cm]{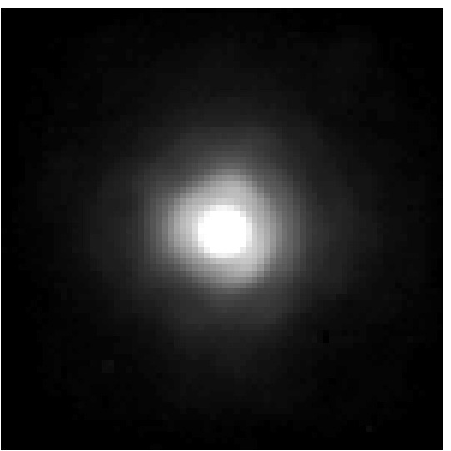}
\caption{NACO images of Achernar A \& B (left) and $\delta$\,Phe (right) in the 2.17\,$\mu$m filter obtained on 22 December 2007. The arrows indicate the position of Achernar B. The field of view is $1\arcsec$, and the grey scale is normalized for both images to the central part of the PSF. \label{k-images}}
\end{figure}

We observed Achernar at several epochs in the second half of 2007 using the Nasmyth Adaptive Optics System (NAOS, Rousset et al.~\cite{rousset03}) of the Very Large Telescope (VLT), coupled to the CONICA infrared camera (Lenzen et al.~\cite{lenzen98}), abbreviated as NACO. Table~\ref{naco_log} gives the list of the observations of Achernar and the standard star $\delta$\,Phe (\object{HD\,9362}). We selected the smallest available pixel scale of $13.26 \pm 0.03$\,mas/pix (Masciadri et al.~\cite{masciadri03}), giving a field of view of 13.6$\arcsec$$\times$13.6$\arcsec$. Due to the brightness of Achernar, we employed narrow-band filters at wavelengths $1.094 \pm 0.015$, $1.644 \pm 0.018$, and $2.166 \pm 0.023\,\mu$m (hereafter abbreviated as 1.09, 1.64, and 2.17) together with a neutral density filter (labeled ``{\tt ND2\_short}"), with a transmission of about 1.5\%. The raw images were processed using the Yorick\footnote{http://yorick.sourceforge.net/} and IRAF\footnote{IRAF is distributed by the NOAO, which are operated by the Association of Universities for Research in Astronomy, Inc., under cooperative agreement with the National Science Foundation.} software packages in a standard way, except that we did not subtract the negligible sky background. Examples of the images of Achernar A \& B and $\delta$\,Phe are presented in Fig.~\ref{k-images}.

\subsubsection{Astrometry}

\begin{figure}[]
\centering
\includegraphics[width=8.3cm]{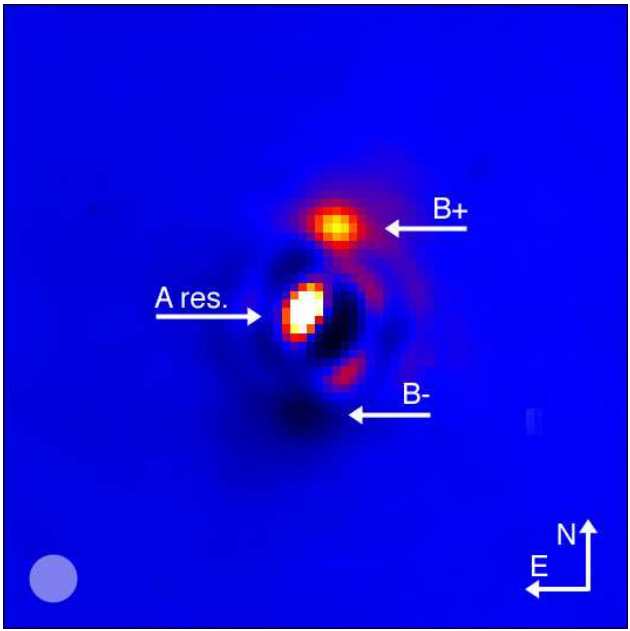}
\caption{Result of the subtraction of a 180$^\circ$ rotated version of the image of the Achernar system presented in Fig.~\ref{k-images} (left) from itself. The ``B+" and ``B-" arrows point at the positions of the positive and negative images of Achernar B, and the ``A res." arrow indicates the A subtraction residuals. The aperture used for photometry is shown in the lower left corner, and the field of view is $1\arcsec$ on each side. \label{image-sub}}
\end{figure}

To obtain the position of B, we first prepared the images by subtracting a 180$^\circ$ rotated version of each image to itself. This removes the contribution of Achernar A and mainly leaves a pair of positive and negative images of B, corrected from the PSF wing leaks from A (Fig.~\ref{image-sub}). We then directly measured the position of Achernar B relative to A using a Gaussian fit procedure on these images. Although Achernar B is clearly visible at all three wavelengths, the Gaussian fit converged essentially in the $K$ band where the high Strehl ratio (compared to the 1.09\,$\mu$m images in particular) results in a better separation of the two objets. The formal fitting error is below $\pm 0.1$\,pixel, but we estimate the true uncertainty to $\pm 0.5$\,pixel (0.066$\arcsec$), due to the presence of residual speckles from Achernar~A close to B. The pixel scale and detector orientation introduce negligible systematic uncertainties (Masciadri et al.~\cite{masciadri03}; Chauvin et al.~\cite{chauvin05}).
%
%
\begin{table}
\caption{Position of Achernar B relative to Achernar A.}
\label{naco_astrom}
\begin{small}
\begin{tabular}{lrccccc}
\hline
Epoch & $\lambda$\,($\mu$m) & $\Delta \alpha$$^{\mathrm{a}}$ & $\sigma(\alpha)$ & $\Delta \delta$$^{\mathrm{a}}$ & $\sigma(\delta)$ & $R_p$$^{\mathrm{b}}$ \\
\hline
2006.760$^{\mathrm{c}}$ & 11.25 & -184.0 & 7.5 & 211.0 & 7.5 & 12.3 \\
2007.488 & 2.17 & -95.9 & 6.6 & 178.6 & 6.6 & 8.9 \\
2007.824 & 2.17 & -46.9 & 6.6 & 165.8 & 6.6 & 7.6 \\
2007.975 & 2.17 & -35.3 & 6.6 & 148.2 & 6.6 & 6.7 \\
\hline
\end{tabular}
\begin{list}{}{}
\item[$^{\mathrm{a}}$] Differential coordinates and uncertainties in milliarcseconds (mas).
\item[$^{\mathrm{b}}$] Projected A--B separation in AU.
\item[$^{\mathrm{c}}$] From Kervella \& Domiciano de Souza~(\cite{kervella07}).
\end{list}
\end{small}
\end{table}

\onltab{1}{ 
\begin{table}
\caption{Log of the NACO observations of Achernar and $\delta$\,Phe.}
\label{naco_log}
\begin{small}
\begin{tabular}{cccccc}
\hline
Date / UT & Star & Filt. & $\Delta t \times N$$^{\mathrm{a}}$ & $\sigma\,(\arcsec)$$^{\mathrm{b}}$ & AM$^{\mathrm{c}}$ \\
\hline
{\it Imaging} \\
2007-06-27T08:48:52 & $\alpha$\,Eri & 2.17 & 2$\times$5 & 0.102 & 1.46 \\
2007-08-01T10:07:20 & $\alpha$\,Eri & 2.17 & 2$\times$5 & 0.082 & 1.19 \\
2007-08-09T09:51:54 & $\alpha$\,Eri & 2.17 & 2$\times$5 & 0.088 & 1.20 \\
2007-10-27T04:41:17 & $\alpha$\,Eri & 2.17 & 2$\times$30 & 0.091 & 1.20 \\
2007-10-27T04:46:27 & $\alpha$\,Eri & 1.64 & 2$\times$22 & 0.092 & 1.20 \\
2007-10-27T04:50:06 & $\alpha$\,Eri & 1.09 & 2$\times$20 & 0.141 & 1.20 \\
2007-10-27T05:45:31 & $\delta$\,Phe & 2.17 & 4$\times$8 & 0.094 & 1.19 \\
2007-10-27T05:48:24 & $\delta$\,Phe & 1.64 & 4$\times$6 & 0.096 & 1.20 \\
2007-10-27T05:50:22 & $\delta$\,Phe & 1.09 & 4$\times$5 & 0.122 & 1.20 \\
2007-10-28T03:01:11 & $\alpha$\,Eri & 2.17 & 1.5$\times$40 & 0.083 & 1.21 \\
2007-10-28T03:04:21 & $\alpha$\,Eri & 1.64 & 1.5$\times$30 & 0.084 & 1.20 \\
2007-10-28T03:07:18 & $\alpha$\,Eri & 1.09 & 1.5$\times$27 & 0.104 & 1.20 \\
2007-10-28T03:37:24 & $\delta$\,Phe & 2.17 & 3$\times$9 & 0.084 & 1.10 \\
2007-10-28T03:39:15 & $\delta$\,Phe & 1.64 & 3$\times$8 & 0.090 & 1.10 \\
2007-10-28T03:44:21 & $\delta$\,Phe & 1.09 & 3$\times$7 & 0.102 & 1.10 \\
2007-11-01T00:09:42 & $\alpha$\,Eri & 2.17 & 0.4$\times$150 & 0.096 & 1.54 \\
2007-11-01T00:12:42 & $\alpha$\,Eri & 1.64 & 0.4$\times$110 & 0.103 & 1.53 \\
2007-11-01T00:15:35 & $\alpha$\,Eri & 1.09 & 0.4$\times$100 & 0.175 & 1.51 \\
2007-11-07T02:22:55 & $\alpha$\,Eri & 2.17 & 0.4$\times$150 & 0.093 & 1.21 \\
2007-11-07T02:25:12 & $\alpha$\,Eri & 1.64 & 0.4$\times$110 & 0.097 & 1.20 \\
2007-11-07T02:28:02 & $\alpha$\,Eri & 1.09 & 0.4$\times$100 & 0.132 & 1.20 \\
2007-11-07T02:37:56 & $\delta$\,Phe & 2.17 & 1$\times$25 & 0.089 & 1.11 \\
2007-11-07T02:39:46 & $\delta$\,Phe & 1.64 & 1$\times$22 & 0.094 & 1.10 \\
2007-11-07T02:41:25 & $\delta$\,Phe & 1.09 & 1$\times$20 & 0.119 & 1.10 \\
2007-12-22T01:39:46 & $\alpha$\,Eri & 2.17 & 1.2$\times$50 & 0.073 & 1.23 \\
2007-12-22T01:44:21 & $\alpha$\,Eri & 1.64 & 1.2$\times$37 & 0.071 & 1.24 \\
2007-12-22T01:46:51 & $\alpha$\,Eri & 1.09 & 1.2$\times$34 & 0.087 & 1.24 \\
2007-12-22T02:26:40 & $\delta$\,Phe & 2.17 & 1$\times$25 & 0.076 & 1.23 \\
2007-12-22T02:31:48 & $\delta$\,Phe & 1.64 & 1$\times$22 & 0.070 & 1.24 \\
2007-12-22T02:34:57 & $\delta$\,Phe & 1.09 & 5$\times$4 & 0.093 & 1.25 \\
\hline
{\it Spectroscopy} \\
2007-10-27T05:22:23 & $\alpha$\,Eri & H & 0.5 $\times$ 100 & 1.28 & 1.23 \\
2007-10-27T05:23:26 & $\alpha$\,Eri & H & 0.5 $\times$ 100 & 1.28 & 1.23 \\
2007-10-27T05:24:20 & $\alpha$\,Eri & H & 0.5 $\times$ 100 & 1.26 & 1.24 \\
2007-10-27T05:25:24 & $\alpha$\,Eri & H & 0.5 $\times$ 100 & 1.23 & 1.24 \\
2007-10-28T03:20:30 & $\alpha$\,Eri & H & 0.5 $\times$ 100 & 1.25 & 1.19 \\
2007-10-28T03:21:33 & $\alpha$\,Eri & H & 0.5 $\times$ 100 & 1.17 & 1.19 \\
2007-10-28T03:22:27 & $\alpha$\,Eri & H & 0.5 $\times$ 100 & 1.20 & 1.19 \\
2007-10-28T03:23:31 & $\alpha$\,Eri & H & 0.5 $\times$ 100 & 1.19 & 1.19 \\
2007-12-22T02:04:11 & $\alpha$\,Eri & H & 1.0 $\times$ 50 & 0.91 & 1.26 \\
2007-12-22T02:05:14 & $\alpha$\,Eri & H & 1.0 $\times$ 50 & 0.91 & 1.26 \\
2007-12-22T02:06:08 & $\alpha$\,Eri & H & 1.0 $\times$ 50 & 0.83 & 1.27 \\
2007-12-22T02:07:10 & $\alpha$\,Eri & H & 1.0 $\times$ 50 & 0.85 & 1.27 \\
\hline
\end{tabular}
\end{small}
\begin{list}{}{}
\item[$^{\mathrm{a}}$] The exposure times $\Delta t$ are given in seconds. $N$ is the number of individual exposures.
\item[$^{\mathrm{b}}$] For the imaging observations, $\sigma\,(\arcsec)$ is the FWHM of the star image as measured on the images themselves, and for the spectroscopic observations, we list the observatory seeing in the visible.
\item[$^{\mathrm{c}}$] AM is the airmass.
\end{list}
\end{table}
} 

\begin{figure}[]
\centering
\includegraphics[width=8.5cm]{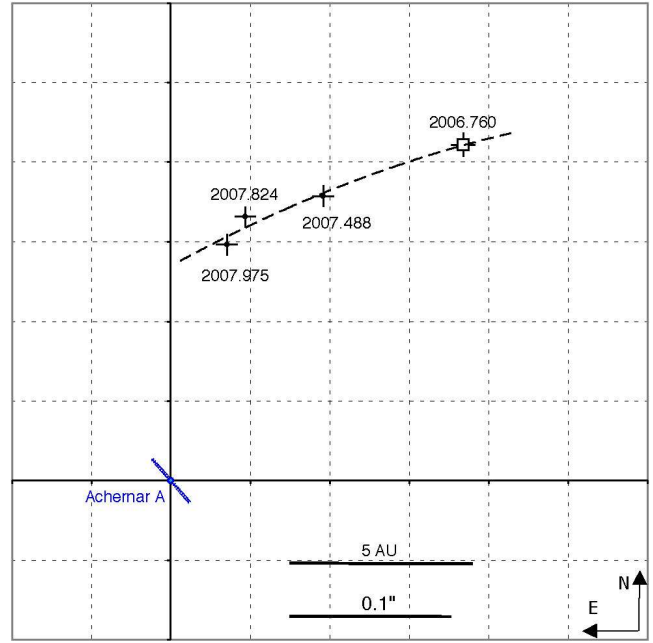}
\caption{Position of Achernar B relative to A for four epochs. The open square indicates the VISIR observation (Kervella \& Domiciano~\cite{kervella07}), and the dots represent the new NACO epochs. The dashed curve is a simple quadratic fit through the data points intended to guide the eye. The segment over Achernar A indicates its projected rotation axis and polar wind as measured by Kervella \& Domiciano~(\cite{kervella06}). The apparent angular sizes of Achernar A and its polar wind are approximately represented to scale. \label{astrometry-fig}}
\end{figure}

Between epochs 2006.760 (Kervella \& Domiciano~\cite{kervella07}) and 2007.975, we measured an apparent displacement of B relative to A of $\rho_B = 161 \pm 10$\,mas along an azimuth of $\alpha = +113^\circ$ (Fig.~\ref{astrometry-fig} and Table~\ref{naco_astrom}).
According to the {\it Hipparcos} catalog (ESA~\cite{esa97}), the amplitude of Achernar A's proper motion on the sky over 1.2\,yr is $\rho_{\rm pm} = 116$\,mas along an azimuth of $\alpha_{\rm pm} = +115^\circ$. If B was a background source, we would expect an apparent displacement \emph{opposite} in azimuth to A's proper motion. As B is clearly comoving with A, we can rule out the possibility that it is a background source. As a remark, the parallactic oscillation of the position of Achernar A is small ($\pi = 22.68 \pm 0.57$\,mas) compared to the observed displacements, but the presence of B could have affected the {\it Hipparcos} proper motion and parallax measurement of Achernar. The other epochs listed in Table~\ref{naco_log} give astrometric positions compatible with the trajectory shown in Fig.~\ref{astrometry-fig}, although with lower accuracy due to poor seeing conditions.

\subsubsection{Photometry}

For the photometric calibration of our images, we observed a standard star immediately before or after Achernar, $\delta$\,Phe (\object{HD\,9362}, G9III) in the same narrow-band filters. It was chosen in the Cohen et al.~(\cite{cohen99}) catalogue of spectrophotometric standards for infrared wavelengths. Due to the small separation of the Achernar A-B pair, PSF fitting is not possible for measuring the photometry of B. We therefore proceeded in three steps:
\begin{enumerate}
\item We measured classical aperture photometry of Achernar relative to $\delta$\,Phe, using a large aperture of 1.2$\arcsec$ in radius.
\item As for the astrometry measurement, we subtracted a 180$^\circ$ rotated version of each image to itself to remove the contribution of Achernar A and leave a positive and a negative image of B,
along with limited residuals from A (Fig.~\ref{image-sub}).
\item We obtained aperture photometry of B from the subtracted images using an aperture radius of 0.040$\arcsec$ (shown in Fig.~\ref{image-sub}).
\end{enumerate}
This procedure has ensured that we obtain properly referenced photometry for Achernar A (the influence of B is negligible at step 1), that we can transfer to B at step 3. Step 2 is necessary because B is located inside the wings of the PSF of A. This subtraction is efficient thanks to the good circular symmetry of the PSF produced by NACO. At step 3, we measure the photometry of A and B over the same aperture and within the same images. This allows us to avoid the problem of the variable Strehl ratio of adaptive optics images.
The conversion of the measured narrow-band magnitudes to standard $JHK$ band magnitudes requires taking the position of the quasi-monochromatic wavelengths within the bands and the shape of the observed spectra into account. Starting from the narrow-band fluxes measured on Achernar and $\delta$\,Phe, we used the Pickles~(\cite{pickles98})\footnote{http://www.ifa.hawaii.edu/users/pickles/AJP/hilib.html} reference spectra corresponding to their spectral types (B3V and G9III) to recover the corresponding broadband flux ratio. The filter profiles were taken from Bessell \& Brett~(\cite{bessell88}). From the October-December 2007 observations, we derived magnitudes of:
$m_J(A) = 0.58_{\pm 0.14}$, $m_H(A)= 0.80_{\pm 0.11}$, $m_K(A) = 0.81_{\pm 0.11}$.
These values generally agree with the $JHK$ magnitudes of Achernar from the 2MASS (Skrutskie et al.~Ê\cite{skrutskie06}; $0.82_{\pm 0.25}$, $0.87_{\pm 0.32}$, $0.88_{\pm 0.33}$) and Ducati~(\cite{ducati02}; $0.79$, $0.86$, $0.88$) catalogs, although systematically brighter by 0.1-0.2\,mag, particularly in the $J$ band. This may be due our narrow-band filters corresponding to the emission lines sometimes present in the spectrum of Achernar.
As we may slightly overestimate the brightness of Achernar, we quadratically added an uncertainty of 0.1\,mag.

From the magnitudes of Achernar A, the small-aperture photometry obtained at step 3 gives the following broadband magnitudes for Achernar B:
$m_J(B) = 4.23_{\pm 0.36}$, $m_H(B)= 4.29_{\pm 0.22}$, $m_K(B) = 4.61_{\pm 0.27}$.
The magnitude differences measured between A and B in the three narrow-band filters are:
$\Delta m_{1.09} = 3.64_{\pm 0.47}$, $\Delta m_{1.64} = 3.50_{\pm 0.19}$, $\Delta m_{2.17} = 3.80_{\pm 0.25}$,
corresponding to an average contrast of $\approx 30$ between the two stars in the near-infrared.
These magnitudes of B are averages over the October-December 2007 observations (epoch 2007.8) and the error bars contain the statistical uncertainty and the dispersion of all the measurements. The magnitudes at the other epochs (list of the observations in Table~\ref{naco_log}) show a marginally significant brightening of B by $-0.7_{\pm 0.4}$\,mag in the $K$ band between 2007.488 and 2007.975 as it approaches A.

\subsection{Spectroscopy}

\begin{figure}[ht]
\centering
\includegraphics[width=8.5cm]{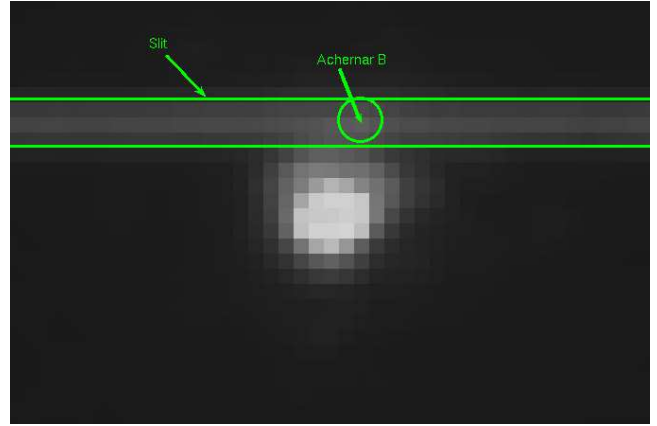}
\caption{Position of the NACO slit relative to Achernar A and B.
} \label{slit-fig}
\end{figure}

\begin{figure*}[ht]
\centering
\includegraphics[width=18cm]{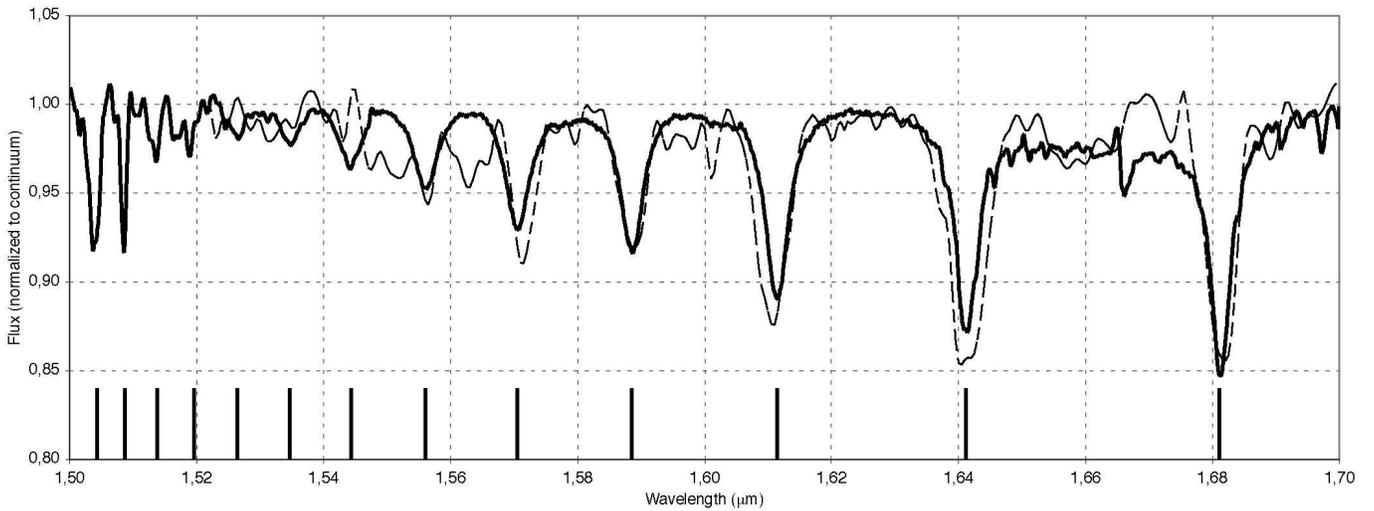}
\caption{Average spectrum of Achernar B (solid curve) compared to the $H$ band spectrum of \object{HR 2763} ($\lambda$\,Gem, A3V, dashed curve) from Ranade et al.~(\cite{ranade04}). The positions of the Brackett transitions 11-4 to 23-4 of hydrogen are marked with vertical lines.} \label{H_spectra-fig}
\end{figure*}

For the spectroscopic observations, we used the {\tt S27\_3\_SH} mode of NACO, featuring a slit width of 86\,mas on the sky, a spectral resolution of 1500 over the $H$ band (dispersion of 0.34\,nm/pixel), and an angular pixel scale of 27\,mas/pixel.
As shown in Fig.~\ref{slit-fig}, the slit was centered on Achernar B. Although most of the light fed into the slit comes from the companion, it is not excluded that Achernar A may contribute with part of the $H$ band spectrum. Details on the 3 recorded ABBA sequences are given in Table~\ref{naco_log}.
Observations were performed using the classical IR NACO sequence where the spectrum is recorded with the target positioned at two different slit positions, i.e. the ABBA observing sequence. For the data reduction we applied two different procedures: (1) the classical spectral reduction procedure using the IRAF packages for dark and flat field corrections, and (2) the A-B, B-A procedure. Because the sky background is negligible and the target is rather bright, both procedures give identical results. In the last step, IRAF was used to obtain 1D spectra from weighted averages of 2D spectra and to perform the wavelength calibration.
Figure~\ref{H_spectra-fig} shows the continuum-normalized average spectrum derived from our 12 observations. The atmospheric absorption lines were corrected using the transmission by Lord~(\cite{lord92}), but some residuals are visible in particular around $\lambda=1.66\,\mu$m (CH$_4$ and water vapor lines).  The absorption lines from the Brackett transitions 11 to 23 of hydrogen are clearly visible in the spectrum and marked by vertical lines.

%

\section{Discussion \label{discussion}}

The {\it Hipparcos} parallax of Achernar ($\pi = 22.68 \pm 0.57$\,mas; ESA~\cite{esa97}) corresponds to a distance modulus of $\mu=3.22 \pm 0.06$\,mag. The absolute magnitudes of B in the $JHK$ and thermal infrared $N$ band (average of the PAH1 and PAH2 magnitudes, see Appendix~\ref{thermal1125}) are therefore:
$M_J = 1.00_{\pm 0.37}$, $M_H = 1.07_{\pm 0.22}$, $M_K = 1.39_{\pm 0.28}$, and $M_N = 1.88_{\pm 0.12}$.
The $JHK$ absolute magnitudes suggest a spectral type around A1V,
slightly fainter than Vega and slightly brighter than Sirius,
which have respective magnitudes of $M_{JHK} \approx 0.55$ and 1.56.
The $N$ band absolute magnitude of Achernar B is also very similar
to that of Sirius ($M_N = 1.62$). It thus does not appear to present
an infrared excess that would betray a Vega-like dusty envelope.
Because Vega itself presents a significant excess in this band
($M_N = 0.55$) due to the presence of dust, it is not comparable.

\begin{figure}[ht]
\centering
\includegraphics[width=8.5cm]{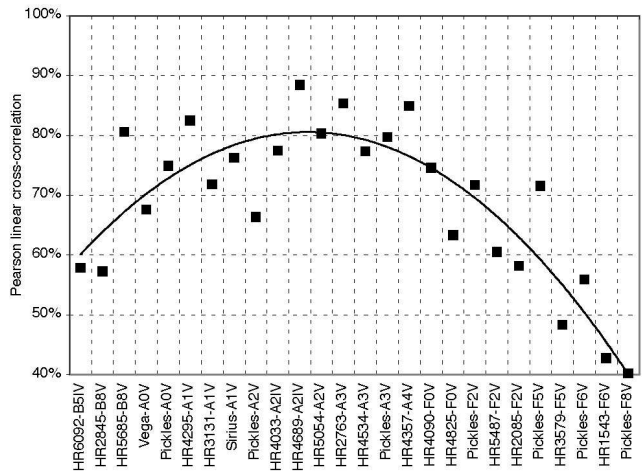}
\caption{
Cross-correlation between our spectrum of B and
reference spectra from Ranade et al.~(\cite{ranade04}),
Meyer et al.~(\cite{meyer98}), and Pickles~(\cite{pickles98}).} \label{Sptype}
\end{figure}
Our NACO $H$ band spectrum of B presents a lot of similarity with early A-type star spectra,
as shown in Fig.~\ref{H_spectra-fig}. This is confirmed by the computation of linear Pearson
cross-correlation coefficients between our normalized spectrum and the reference $H$ band spectra
from Ranade et al.~(\cite{ranade04})\footnote{http://vo.iucaa.ernet.in/$\sim$voi/NIR{\textunderscore}Header.html} and Pickles~(\cite{pickles98})\footnote{http://www.ifa.hawaii.edu/users/pickles/AJP/hilib.html}.
As shown in Fig.~\ref{Sptype}, the spectral types A0V-A4V give the best correlation with our observation.
As shown by Meyer et al.~(\cite{meyer98})\footnote{ftp://ftp.noao.edu/catalogs/medresIR/},
the equivalent width (EW) of the 11-4 Brackett line at $1.681\,\mu$m offers a possibility to test the
effective temperature, although it is not monotonic. An integration normalized to the pseudo-continuum
between 1.670 and $1.690\,\mu$m gives an equivalent width of 0.675\,nm (=2.4\,cm$^{-1}$).
From Fig.~7 in Meyer et al.~(\cite{meyer98}), this gives $\log T_{\rm eff} \approx 3.9-4.1$,
also compatible with an early A-type star.
Such a star has an approximate mass of $\approx 2$\,M$_{\odot}$ (Kervella et al.~\cite{kervella03}).

Although we cannot derive the full parameters of Achernar~B's orbit from our limited astrometry,
the combination of its estimated mass ($\approx 2$\,M$_\odot$), the mass of A ($\approx 6.7$\,M$_\odot$; 
Vinicius et al.~\cite{vinicius06}, see also Harmanec~\cite{harmanec88})
and the maximum apparent A-B separation ($r_{AB} \approx 12.3$\,AU)
allow us to roughly estimate its period. 
From Kepler's third law and by assuming an elliptic orbit, we have $T^2 = a^3/M$ where $a$ is the semi-major axis in AU,
$T$ the period in years, and $M$ the total mass in M$_\odot$. Assuming that $a$ is equal to the observed maximum
separation of 12.3\,AU, we obtain a minimum period of $T = 14$ to 15\,yr.

\section{Conclusion}

From our photometry and spectroscopy, Achernar B is most probably an A1V-A3V star.
Our data are currently insufficient for deriving its full orbit, but its minimum period is $\approx 15$\,yr.
Its fast orbital motion should allow the derivation of reliable parameters within a few years.
The periodic approach of the companion could be the cause of the observed $\approx 15$\,yr pseudo-periodicity of the Be episodes of Achernar (Vinicius et al.~\cite{vinicius06}). The passage of B at periastron within a few AUs of A could
extract material from the equator of A, where the effective gravity is very low.
The next such passage should happen around 2010 (Meilland~\cite{meilland07}).
Achernar appears similar to the B0.2IVe star $\delta$\,Sco, which
has a 1.5\,mag fainter companion on a highly excentric 10.6\,yr orbit (Bedding~\cite{bedding93};
Miroshnichenko et al.~{\cite{miro01}).
This suggests that the presence of companions around Be
stars should be examined carefully, as it may play a key role in triggering the Be phenomenon.

\begin{acknowledgements}
This research made use of the SIMBAD and VIZIER databases at the CDS, Strasbourg (France),
and NASA's Astrophysics Data System Bibliographic Services. We also received the support of PHASE, the high angular resolution partnership between ONERA, Observatoire de Paris, the CNRS, and University Denis Diderot Paris 7.
\end{acknowledgements}


 \Online
\appendix

\section{Thermal infrared imaging at 11.25\,$\mu$m \label{thermal1125}}

Kervella \& Domiciano de Souza~(\cite{kervella07}) did not detect Achernar~B in the 11.25\,$\mu$m images (PAH2 band) of the system obtained with the VLT/VISIR instrument in BURST mode. A re-analysis of these observations has shown that the reason for this non-detection was the introduction of noise during the normalization and subtraction of the PSF reference star ($\delta$\,Phe).
For our VISIR program on the B[e] star MWC300 (Domiciano de Souza et al.~\cite{domiciano08}), we obtained BURST mode images of another PSF calibrator ($\eta$\,Ser) 3.4\,hours before the observations of Achernar (under similar seeing conditions). Since $\eta$\,Ser is brighter than $\delta$\,Phe, we could obtain a cleaner PSF-subtracted image of Achernar~B, as presented in Fig.~\ref{VisirPAH2}. From this image, aperture photometry over a 0.22$\arcsec$ diameter gives a flux ratio of 1.74\% between Achernar B and A (the ratio of peak intensities is 2.14\%). From the 16.8\,Jy absolute flux derived by Kervella \& Domiciano de Souza~(\cite{kervella07}) for A, the contribution from B is therefore 0.3\,Jy at 11.25\,$\mu$m. This value is comparable to the 0.4\,Jy flux derived by Kervella \& Domiciano de Souza~(\cite{kervella07}) at 8.59\,$\mu$m.

\begin{figure}[]
\centering
\includegraphics[width=8.9cm]{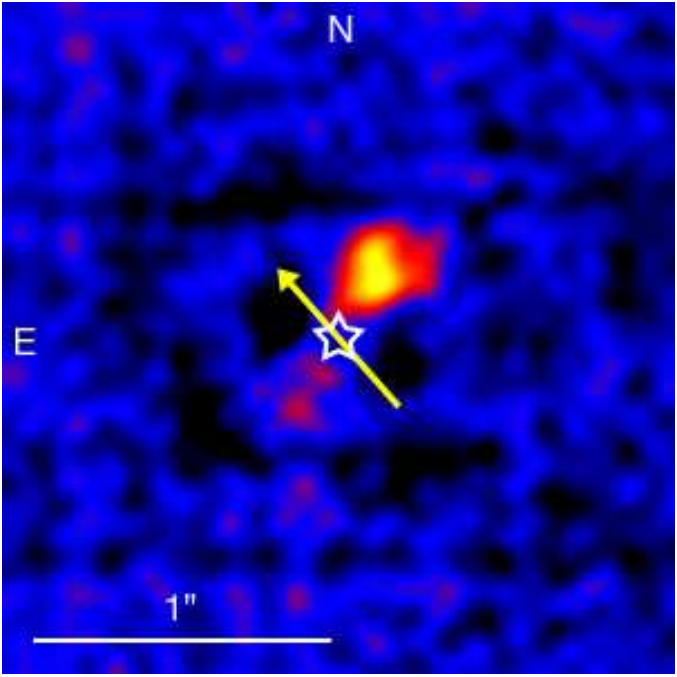}
\caption{Residual of the subtraction of the VISIR PAH2 image of $\eta$\,Ser
from the image of Achernar in the same band. The contribution of Achernar~B is clearly
visible. The arrow indicates the rotation axis of Achernar~A. \label{VisirPAH2}}
\end{figure}

\end{document}